\documentstyle[12pt]{article}
\begin{document}
\title{Issues in Quantized Fractal Space Time}
\author{B.G. Sidharth\\Centre for Applicable Mathematics \& Computer Sciences\\
B.M. Birla Science Centre, Hyderabad 500 063}
\date{}
\maketitle
\begin{abstract}
In recent years, the picture of discrete space time has been studied in the
context of stochastic theory. There are a number of ramifications, which are
briefly examined.
We argue that the causality of physiics has its roots in the analyticity within
the two dimensons of a fractal quantum path and further show how this picture
has convergence wth quantum superstrings.
\end{abstract}
\section{Introduction}
The concept of a discrete space time was formulated over fifty years ago by
Schild, Snyder and others\cite{r1,r2}. It has survived over the years with
variations through the works of Caldirola, Lee, Bombelli et al., Wolfe,
the author and others (\cite{r3}-\cite{r11}). Infact, space time points imply
infinite energies and momenta, but Quantum Theory has lived with this self
contradiction\cite{r12}. Devices like renormalization have therefore to be
invoked.\\
A second approach which avoids the difficulties of both the classical smooth
and the discrete theory is that pioneered by Ord, Nottalee and El Naschie\cite{r54}-
\cite{r56}. Those authors assume the existence of a transfinite Cantor-like space
time Manifold $\mathcal{E}^{(\infty)}.$\\
This concept is also implicit in Dirac's relativistic theory of the electron
\cite{r13}: Physically meaningful solutions arise only at and above the Compton
scale. Below it we have the zitterbewegung effects.\\
Recently the author has pointed out that the electron can be fruitfully described
with a Kerr-Newman metric, avoiding a naked singularity by invoking a discrete
space time structure, which can be further pushed forward in the context of a
stochastic and fractal theory\cite{r14}-\cite{r18}. Such a
formulation leads to a unification of electromagnetism and gravitation, and
also shows up a cosmology consistent with the large number relations and recent
observations, apart from providing a rationale for many other apparently ad hoc
features. We will now briefly examine the above formulation and some of its
ramifications.
\section{The Kerr-Newman Formulation}
It is well known that the Kerr-Newman metric describes the field of an electron
including the anomalous gyro magnetic ratio $g=2$ \cite{r19} except that there is
an inadmissable naked singularity: The radius of the horizon for the electron
is given, in the usual notation by
\begin{equation}
r_+ = \frac{GM}{c^2} + \imath b, b \equiv \left(\frac{G^2Q^2}{c^8} + a^2
- \frac{G^2M^2}{c^4}\right)^{1/2}\label{e1}
\end{equation}
It is to be noticed that the imaginary part of the radius in the above equation
can be immediately identified with the imaginary part of the position operator
of a Dirac electron, viz.,
\begin{equation}
x = (c^2p_1H^{-1}t + a_1) + \frac{\imath}{2} c\hbar (\alpha_1 - cp_1H^{-1})H^{-1},\label{e2}
\end{equation}
The imaginary part in (\ref{e2}) gives the zitterbewegung and vanishes on averaging
over Compton scales. It has been stressed\cite{r14,r15} that the Compton scale
represents a minimum space time cut off, below which there is no meaningful
physics. Physics emerges on averaging over these scales.\\
These minimum cut off scales have been shown to arise due to a stochastic underpinning
\cite{r20,r21}. Indeed they give a meaning to Nelsonian and other stochastic studies
\cite{r22}-\cite{r25}. All this is symptomatic of a background Zero Point
Field or Quantum Vacuum which is the underpinning for the universe.\\
The above Kerr-Newman formulation of the electron becomes more meaningful in a
linearised General Relativistic context, where we have,
\begin{equation}
g_{\mu v} = \eta_{\mu v} + h_{\mu v},h_{\mu v} = \int \frac{4T_{\mu v}(t-|\vec x -
\vec x'|, \vec x')}{|\vec x - \vec x'|}d^3x'\label{e3}
\end{equation}
The justification for the linearized theory is that even at the Compton scale
of the electron $\sim 10^{-11}cms$ we are well outside the Schwarzchild radius
of an electron $\sim 10^{-56}cms$\cite{r26} so that the linearized theory
is applicable. Starting from (\ref{e3}) it has been
shown that the correct spin, charge, gyro magnetic ratio, and infact the Kerr
Newman metric itself can be obtained. This metric gives both the gravitational
and the electromagnetic fields, and it has been shown that this is symptomatic
of and leads to the desired
unification of gravitation and electromagnetism. The formulation is on the face
of it similar to Weyl's original theory, except that what is crucial is the purely
Quantum Mechanical spinorial behaviour of the electron. Indeed the
electromagnetic potential is given by
$$A^\mu = \hbar \Gamma^{\mu \sigma}_{\sigma} = \hbar \frac{\partial}{\partial x^\mu}
log(\sqrt{|g|})$$
which is apparently similar to the Weyl theory, except that this time the
term is not put in ad hoc as in the Weyl formulation, but rather arises naturally
due to the pseudo spinorial behaviour of the negative energy two component
spinor $\chi$ of the Dirac four spinor $\left(\begin{array}{c} \Theta \\ \chi
\end{array}\right)$. All this has been discussed
in detail in the references\cite{r14,r15,r16}.\\
Further at the Compton
scale one gets a characterisation of the quark picture including such peculiar
features as the fractional charge, handedness and confinement\cite{r18,r27,r20}.
A crucial input here is the fact that while the usual three dimensionality
arises well outside the Compton scale due to spin networks\cite{r19,r28}, at
the Compton scale itself we encounter lower dimensionality. Indeed, in the
closely related non-Quantum Mechanical relativistic theory of particles \cite{r29},
the centres of mass form a two dimensional disc within the analogue of the
Compton scale.\\
We also get a characterisation of the neutrino and weak interaction\cite{r30}.
Indeed in the above model the Compton wavelength of the
neutrino becomes infinite as its mass vanishes and so the divide
between the negative energy spinors and the positive energy spinors of the four
spinor as in the case of the electron disappears, and along with it the double
connectivitity of space also. This immediately leads to the handedness of the neutrino
\cite{r14}. Infact the neutrino is the divide between the Fermionic and
the Bosonic particle.
\section{Cosmology}
It is easy to see how the Kerr-Newman type electrons, or more generally
elementary particles can be formed out of a background Zero Point Field. As
is known the energy of the fluctuations of the magnetic field in a region of
length $\lambda$ is\cite{r19}
\begin{equation}
B^2 \sim \frac{\hbar c}{\lambda^4}\label{e4}
\end{equation}
where $B$ to the left side of (\ref{e4}) denotes the magnitude of the magnetic
field. If $\lambda$ is the Compton wavelength, then the right side $\sim mc^2$, that
is the whole particle can be created out of the background Zero Point Field. In
what follows we take a pion to be a typical elementary particle, as in the literature,
its mass being $m$. As there are $N \sim 10^{80}$ particles in the universe,
we should then have
\begin{equation}
Nm = M,\label{e5}
\end{equation}
where $M \sim 10^{56}gms$ is the mass of the universe. This is indeed so.\\
We can next deduce using the ZPF spectral density, the relation (cf.ref.\cite{r15}),
$M \alpha R$ where $R$ is the radius of the universe. This is quite correct and infact poses a puzzle, as is
well known and it is to resolve this dependence that dark matter has
been postulated whereas in our formulation the correct mass radius
dependence has emerged quite naturally.
Other interesting and consistent  consequences follow from the facts that
the pion energy equals its gravitational energy and that
$\sqrt{N}$ particles are fluctuationally created in Compton time $\tau$:
\begin{equation}
\frac{GM}{c^2} = R, \sqrt{N} = \frac{2m_\pi c^2}{\hbar} .T\label{e6}
\end{equation}
where $T$ is the age of the universe $\approx 10^{17}secs$, and,
\begin{equation}
H = \frac{Gm_\pi^3 c}{\hbar^2},\label{e7}
\end{equation}
It is remarkable that equation (\ref{e7}) is known to be true from a
purely empirical standpoint and is considered mysterious. Remembering that we
are dealing with order of magnitude relations, we can deduce from
(\ref{e6}) and (\ref{e7}) that
\begin{equation}
\frac{d^2R}{dt^2} =\Lambda R \quad \mbox{where}\quad\Lambda \leq 0 (H^2)\label{e8}
\end{equation}
That is, a small cosmological constant cannot be ruled out.
Equation (\ref{e8}) explains the puzzling fact that, if $\Lambda$ exists, why is
it so small\cite{r31}.
To proceed we observe that the fluctuation of $\sim \sqrt{N}$ in the
number of particles leads to
$$\sqrt{N} = \frac{e^2}{Gm^2} \approx 10^{40},$$
whence we get,
\begin{equation}
R = \sqrt{N}l, \frac{Gm}{lc^2} = \frac{1}{\sqrt{N}}, G \propto T^{-1},\label{e9}
\end{equation}
as also the fact that $\dot G/G \sim \frac{1}{T}$, in reasonably good agreement
\cite{r32}.\\
Further, from the above we can deduce that the charge $e$ is independant
of time or $N$.
Infact we can treat $m$ (or $l$), $c \mbox{and}\hbar$ as the only microphysical
constants and $N$ as the only cosmological parameter, given which all other
parameters and constants follow.\\
We can now easily deduce from (\ref{e6}), (\ref{e7}) and (\ref{e8}), the following:
\begin{equation}
\rho \propto T^{-1}, \Lambda \leq 0 (T^{-2}),\label{e10}
\end{equation}
In our model the equation (\ref{e6}) actually provides an
arrow of time, in terms of the particle number $N$.
Further, the cosmic background radiation can be explained in terms of
fluctuations of interstellar Hydrogen\cite{r33}. In this model the universe
continues to expand for ever with according to (\ref{e10}) decreasing
density (unlike in the Steady State model).
Indeed the fact that the universe would continue to expand for ever has since
been observationally confirmed\cite{r34,r35}. Moreover, using (\ref{e9})
we can deduce the well known effects of General Relativity namely the precession
of the perihelion of mercury, the gravitational bending of light, a recently
observed anomalous inward acceleration in the solar system and the flattened
galactic rotational curves without invoking dark matter\cite{r36,r30}. It may also be remarked that a background Zero Point Field does
indeed show up as a cosmological constant, as has been shown in (Cf.ref.\cite{r30}).
\section{Other Consequences}
\subsection{Weak Interactions}
These can be characterised in terms of the semionic or anomalous statistics
of the neutrino, as noted in Section 2. Using this we can deduce that
\begin{equation}
\frac{m_\nu c^2}{k} \approx \sqrt{3}T\label{e11}
\end{equation}
At the present background temperature of about $2^\circ K$, this
gives a neutrino mass
\begin{equation}
10^{-9}m_e \le m_\nu \le 10^{-8}m_e\label{e12}
\end{equation}
where $m_e$ is the electron rest mass.
It is remarkable that (\ref{e11}) is exactly what is required to be deduced
theoretically to justify recent models of lepton conservation or in certain
unification schemes.
We now observe that the balance of the gravitational force
and the Fermi energy of the cold background neutrinos, gives\cite{r37}
$$\frac{GN_\nu m_\nu^2}{R} = \frac{N_\nu^{2/3}\hbar^2}{m_\nu R^2},$$
whence, $N_\nu \sim 10^{90}$\\
where $N_\nu$ is the number of neutrinos, as indeed is known.\\
If the new weak force is mediated by an intermediate particle of mass $M$ and
Compton wavelength $L$, we will get from the fluctuation of the particle
number $N_\nu$, on using (\ref{e12}),
\begin{equation}
g^2\sqrt{N_\nu} L^2 \approx m_\nu c^2 \sim 10^{-14},\label{e13}
\end{equation}
From (\ref{e13}), on using the value of $N_\nu$, we get,$g^2L^2 \sim 10^{-59}$\\
This agrees with experiment and the theory of massless particles the neutrino
specifically acquiring mass due to interaction\cite{r38}, using the usual
value of $M \sim 100 Gev.$\\
Additionally there could be a long range force also, a "weak electromagnetism"
with coupling   $\bar g$. This time, in place of (\ref{e13}), we would have,
\begin{equation}
\frac{\bar g^2 \sqrt{N_\nu}}{R} \approx 10^{-8}m_\nu c^2\label{e14}
\end{equation}
Comparing (\ref{e14}) with a similar equation for the electron, we get\cite{r30},
$$\bar g^2/e^2 \sim 10^{-13}$$
so that in effect the neutrino will appear with an "electric charge" a little less than
a millionth that of the electron.\\
Interestingly from an alternative perspective \cite{r39}, it
can be concluded that the cosmological Neutrino background can mediate long
range forces $\sim \frac{1}{r^2}$ for $r << \frac{1}{T}$ and $\sim \frac{1}{r^6}$
for $r > > \frac{1}{T}$.\\
In any case using the value of the Neutrino "electrical"
charge and treating it like a Kerr-Newman black hole, as in the case of the
electron it is easy to deduce a magnetic moment for it:\\
We start with the relation\cite{r15}
$$\hbar \sim \frac{Gm^2\sqrt{N}}{c}$$,
where $m$ is the electron mass. For the neutrino mass as given in (\ref{e12})
and particle number $N_\nu$, the above becomes,
$$\hbar' \sim 10^{-12} \hbar,$$
which is symptomatic of the bosonic (or spin zero) behaviour of the neutrino.\\
Using this value $\hbar'$ instead of $\hbar$, and the neutrino "electric" charge
and mass as given above, we can deduce that its magnetic moment is given
by
$$\mu_\nu \sim 10^{-11} \quad \mbox{Bohr}\quad \mbox{magnetrons}$$
Indeed, this is in agreement with known values \cite{r40}.
\subsection{Discrete Space-time Effects}
Within the above scheme, the neutral pion has been exhibited as a bound state
of an electron and a positron. At first sight one would expect that such a bound
state would cause pair annihilation and would lead to the appearance
of two photons. However the existence of such a bound state is an imprint of
discrete space time as can be seen by the following argument: In this case the
the Schrodinger equation is given by (Cf. also \cite{r4})
\begin{equation}
H \phi T = E \phi T = \phi \imath h [\frac{T(t+\tau/2) - T (t-\tau/2)}{\tau}]
\label{e16}
\end{equation}
where $\phi$ is the space part and $T$ is the time part of the wave function
and $\tau$ is the minimum unit, the Compton time. From (\ref{e16}) one can
deduce that the most excited stable state appears at the critical value
$$E \sim \frac{\hbar}{\tau} \sim mc^2$$
which is ofcourse the pion energy. It must be pointed out that the decay
mode of the pion bears out these arguments.\\
Thus the existence of the pion as a bound state of an electron and a poistron due
to discrete space time is similar to the original Bohr orbits, at the birth of Quantum
Mechanics, as will be seen below.\\
In the same vein, it was argued that the Kaon decay puzzle, wherein time
reversibility is violated, could be explained on similar lines\cite{r41}. Let
us now consider the effect more closely. Indeed it has been shown that
the discreteness leads to a non commutative geometry \cite{r20} viz.,
\begin{equation}
[x,y] = 0 (l^2), [x,p_x]=\imath \hbar [1-l^2]\label{dis1}
\end{equation}
and similar equations. If terms $\sim l^2$ are neglected we get back the usual
Quantum Theory. However if we retain these terms, then we can deduce the
Dirac equation. Moreover it can be seen that given (\ref{dis1}) space
rerlection symmetry no longer holds. This violation is an $O(l^2)$ effect.\\
This is not surprising. It has already been pointed out \cite{r42} that the
space time divide viz., $x + \imath ct$ arises due to the zitterbewegung
or double Weiner process in the Compton wavelength - and in this derivation
terms $\sim (ct)^2 \sim l^2$ were neglected. However if these terms are
retained, then we get a correction to the usual theory including special
relativity.\\
To see this more clearly let us (Cf.ref.\cite{r14,r43})
as a first approximation treat the continuum as a series of discrete points
separated by a distance $l$, which then leads to
\begin{equation}
Ea(x_n) = E_0a(x_n) - Aa(x_n + l) - Aa(x_n - l)\label{dis2}
\end{equation}
When $l$ is made to tend to zero, it was shown that from (\ref{dis2}) we recover the
Schrodinger equation, and further, we have,
\begin{equation}
E = E_0 - 2A cos kl.\label{dis3}
\end{equation}
The zero of energy was chosen such that $E = 2A = mc^2$, the rest energy of the
particle (Cf.\cite{r14}), in the limit $l \to 0$. However if we retain terms
$\sim l^2$, then from (\ref{dis3}) we will have instead
$$\left| \frac{E}{mc^2} - 1\right| \sim 0 (l^2)$$
The above shows the correction to the energy mass formula, where again we
recover the usual formula in the limit $O(l^2) \approx 0$.\\
It must be mentioned that all this would be true in principle for discrete
space time, even if the minimum cut off was not at the Compton scale.\\
Intuitively this should be obvious: Space time reflection symmetries are
based on a space time continuum picture.
\subsection{A Mass Spectrum}
It is possible to obtain the masses of different elementary particles by considering
them to be suitable bound states of the leptons, on the basis of their decay
modes, in the context of the above picture\cite{r44}.\\
For example a proton could be considered to be a bound state of two positrons
and a central electron\cite{r17,r18}. In this case we recover the correct
mass of the proton, $m_P$. There is also a string of excited states with
masses $(2n+1)m_P$. It is quite remarkable that the $\Omega_c$ baryon has a
mass nearly $3 m_P$, after which there is a big gap and the next baryon viz.,
$\Lambda_v$ has a mass $5 m_P$. Similarly other shortlived baryons whose
masses are odd multiples of the proton masses can be expected.\\
Taking into account the rotational degrees of freedom, it is possible to
get an additional mass spectrum(Cf.ref.\cite{r44}) viz.,
$$mc^2 = \frac{2(n+1/2)\hbar \omega}{1-\frac{5}{6}k^2},$$
where for $n=0$ and $k=0$, $\hbar \omega = m_Pc^2$. For $k=1$ this then gives
a mass six times that of a proton and so on we can generate a series of
masses.
\subsection{The Magnetic Effects}
If the electron is indeed a Kerr-Newman type charged black hole, it can be
approximated by a solenoid and we could expect an Aharonov-Bohm type of
effect, due to the vector potential $\vec A$ which would give rise to shift
in the phase in a two slit experiment for example \cite{r45}. This shift
is given by
\begin{equation}
\Delta \delta_{\hat B} = \frac{e}{\hbar} \oint \vec A . \vec{ds}\label{e17}
\end{equation}
while the shift due to the electric charge would be
\begin{equation}
\Delta \delta_{\hat E} = -\frac{e}{\hbar} \int A_0 dt\label{e18}
\end{equation}
where $A_0$ is the electrostatic potential. In the above formulation (Cf.ref.\cite{r46},.
we would have
\begin{equation}
\vec A \sim \frac{1}{c} A_0\label{e19}
\end{equation}
Substitution of (\ref{e19}) in (\ref{e17}) and (\ref{e18}) shows that
the magnetic effect $\sim \frac{v}{c}$ times the electric effect.\\
Further, the magnetic component of a Kerr-Newman black hole, as is well known
\cite{r19} is given by
\begin{equation}
B_{\hat r} = \frac{2ea}{r^3} cos \Theta + 0(\frac{1}{r^4}), B_{\hat \Theta} =
\frac{ea sin\Theta}{r^3} + 0(\frac{1}{r^4}), B_{\hat \phi} = 0,\label{e20}
\end{equation}
while the electrical part is
\begin{equation}
E_{\hat r} = \frac{e}{r^2} + 0(\frac{1}{r^3}), E_{\hat \Theta} = 0(\frac{1}{r^4}),
E_{\hat \phi} = 0,\label{e21}
\end{equation}
Equations (\ref{e20}) and (\ref{e21}) show that in addition to the usual
dipole magnetic field, there is a shorter range magnetic field given by terms
$\sim \frac{1}{r^4}$. In this context it is interesting to note that an extra
$B^{(3)}$ magnetic field of shorter range and mediated by massive photons
has indeed been observed and studied over the past few years\cite{r47}.
\subsection{Unification of Fluctuations}
It is quite remarkable that in the preceding considerations, we get the
gravitational and electromagnetic interactions as also the weak interaction
from fluctuation in the particle number $\sim \sqrt{N_\nu}$ or $\sqrt{N}$, for
example in the equations leading to (\ref{e9}) or (\ref{e13}). It is thus
the fluctuation or what has been characterised as non local effects that
underlie all the interactions\cite{r48} and lead to a unified
picture. It is interesting that El Naschie's fluctuon is very much in this
spirit\cite{r49}.
\subsection{Fractal Matter}
It has been remarked earlier that as we approach the Compton wavelength,
there is a change in dimensionality - we go to two and one dimensions. Such a
low dimensional behaviour leads to fractional charges and superconductivity
type effects\cite{r50,r51,r52}. It is indeed pleasing that experiments with
carbon nano tubes already indicate such phenomenon. Thiis type of behaviour
should be exhibited by quarks also.
\subsection{Miscellaneous Matters}
a) The discrete space time or zitterbewegung has an underpinning that is
stochastic. The picture leads to the goal of Wheeler's 'law without law' \cite{r53,r20,r21}.
Furthermore the picture that emerges is machian. This is evident
from equations like (\ref{e6}), (\ref{e7}) and (\ref{e9})-- the micro depends
on the macro. So the final picture that emerges is on of stochastic holism.\\
b) Another way of expressing the above point is by observing that the
interactions are relational. For example, in the equation leading to (\ref{e9}),
if the number of particles in the universe tends to $1$, then the
gravitational and electromagnetic interactions would be equal, this happening
at the Planck scale, where the Compton wavelength equals the Schwarzchild
radius\cite{r54,r55}.\\
c) Infact as shown \cite{r42}, when $N$ the number of
particles in the universe is $1$ we have a Planck particle with a
short life time $\sim 10^{-42}secs$ due to the Hawking radiation but with
$N \sim 10^{80}$ particles as in the present universe we have the pion as
the typical particle with a stable life time $\sim$ of the age of the
universe due to the Hagedorn on radiation.\\
d) It is well known that there are 18 arbitrary parameters in contemporary
physics. We on the other hand have been working with the micro physical
constants referred to earlier viz., the electron (or pion) mass or Compton
wavelength, the Planck constant, the fundamental unit of charge and the velocity
of light. These along with the number of particles $N$ as the only free parameter
can generate the mass, radius and age of the universe as also the Hubble
constant.\\
If we closely look at the equation leading to (\ref{e9}) giving the gravitational
and electromagnetic strength ratios, we can actually deduce the relation,
\begin{equation}
l = \frac{e^2}{mc^2}\label{e24}
\end{equation}
In other words we have deduced the pion mass in terms of the electron mass, or,
given the pion mass and the electron mass, we have deduced the fine structure
constant. From the point of view of the order of magnitude theory in which
the distinction between the electron, pion and proton gets blurred, what
equation (\ref{e24}) means is, that the Planck constant itself depends on
$e$ and $c$ (and $m$). Further in the Kerr-Newman type characterisation of
the electron, the charge $e$ is really equivalent to the spinorial tensor
density $(N=1)$, Cf.ref.\cite{r14}. In this sense $e$ also is pre determined
and we are left with a minimum length viz. the Compton length and a minumum time
viz. the Compton time (or a maximal
velocity $c$) as the only fundamental constants.

{\bf \Large APPENDIX}\\ \\
{\bf \large A Brief Note on Analyticity and Causality, and the "Levels of
Physics"}\\ \\
In previous communications\cite{r1,r2} it was argued that within the Compton
wavelength of a fractal Brownian path the one dimensional coordinate $x$ becomes
complex. That is $x$ becomes $x+\imath x'$ where, further it was shown that
$x' = ct,$ and it was argued that this is the origin of Special Relativity.
It is relevant to mention that complex time has also been considered by El
Naschie\cite{r3,r4}.\\
Within the Compton scale we have a Hawking-Hartle\cite{r5} situation, where
time becomes imaginary and static - in these regions space time can be
represented by the compact rotation group. This is also from the point of
view of ordinary space-time, the unphysical
zitterbewegung region, and as Dirac\cite{r6} pointed out, physics begins
after an integration over this region. Outside the Compton scale however,
which is our physical domain\cite{r7}, we have the Minkowski metric.\\
Interestingly it was argued\cite{r1} that the above consideration was also at the root of the
complex wave function of Quantum Mechanics, which differenciates it from
classical physics. Infact if we take the Quantum Mechanical wave function to be
the carrier of as much physical information of the state as is possible then we
can argue that because the first time derivative of the wave function is then
not required, unlike in classical theory, where position and velocity are
independantly required, it is necessary that the wave
function be complex in order to preserve causality\cite{r8}. Indeed if the wave
function were real, we would have a stationary picture with a constant
probability current.\\
Within the Compton scale, that is in the domain of the Hawking-Hartle static
time $t'$, we have
\begin{equation}
x^2 + y^2 + z^2 + c^2 + t'^2 = \mbox{invariant}\label{e1}
\end{equation}
Further, analyticity demands the Cauchy-Reimann equations which lead to, in
this case, the Laplacian operator equation,
\begin{equation}
\left[\nabla^2 + \frac{1}{c^2} \frac{\partial^2}{\partial t'^2}\right] \psi = 0\label{e2}
\end{equation}
However, when we cross over to the domain of our usual physics, firstly $t'$ goes
over to $\imath t$, where $t$ is our physical time. Secondly the region of
analyticity in our physical world is the region outside the Compton wavelength,
and excludes the region within which is the unphysical
zitterbewegung region of non local or superluminal effects (Cf.\cite{r6}). So as
Dirac pointed out, after the integration over the
unphysical Compton scale, and remembering that,
$$\psi = \sqrt{\rho}e^{\imath S},$$
where $\rho$ is the probability density of $x(t)$, so that an integration of
$\psi \psi^*$ over the Compton region gives the mass, we get,
instead of (\ref{e1}) and (\ref{e2}),
\begin{equation}
x^2 + y^2 + z^2 - c^2 t^2 = \mbox{invariant}\label{e3}
\end{equation}
\begin{equation}
\Box \psi = mc^2\psi\label{e4}
\end{equation}
Equation (\ref{e3}) is the Minkowski metric while equation (\ref{e4})
can be easily recognised as the Klein-Gordon equation.\\
There is another way to derive the D'Alembertian in (\ref{e4}). For this we use
(\ref{e3}), along with the fact that, precisely due to the double Weiner process
at the Compton scale,
referred to above, as pointed out by Nottale\cite{r9} the energy momentum
operators are given by
$$\left(\frac{\hbar}{\imath} \vec \nabla , \frac{\hbar}{\imath} \frac{\partial}
{\partial t}\right)$$
Thus causality in the physical world as expressed by (\ref{e3}) or (\ref{e4})
is related to analyticity within the unphysical Compton region.\\
We next observe that as above, the discretization at the Compton scale leads to
the commutation relations\cite{r10,r11}:
\begin{equation}
[x,p_x] = \imath \hbar [1-l^2]\label{e5}
\end{equation}
Equation (\ref{e5}) implies a correction to the usual uncertainity relation,
due to the minimum space-time cut off. This is exactly the case in Quantum
superstrings\cite{r12,r13}. There also, due to duality, a minimum cut off emerges
and we have the equation (\ref{e5}).\\
What we would like to point out is that we are seeing here different levels of
physics. Indeed, rewriting (\ref{e5}) as,
$$[x,u_x] = \imath [l-l^3],$$
we can see that if $l=0$, we have classical physics, while if $0(l^3)=0,$ we have
Quantum Mechanics and finally if $0(l^3)\ne 0$ we have the above discussed
fractal picture, and from another point of view, the superstring picture.\\
Interestingly, in our case the electron Compton wavelength $l \sim 10^{11}cm,$ so
that $0(l^3)\sim 10^{-33}$ as in string theory.

\end{document}